\documentclass[12pt,a4paper]{article}
\usepackage{graphics}

\begin{document}
\thispagestyle{empty}
\vspace*{1cm}

\begin{flushright}
{\bf ISU-IAP.Th 2001-05, Irkutsk}
\end{flushright}
\vspace*{1cm}

\begin{center}
{\Large {\bf Unitary mixing scalar--vector in $\xi$ gauge}
\\[1.0cm]
A.E.Kaloshin \footnote{EM:\ kaloshin@physdep.isu.runnet.ru },\ \ A.E.Radzhabov} \\[1cm]
{\sl Institute of Applied Physics, Irkutsk State University\\
blvd Gagarin 20, 664003 Irkutsk, Russia and \\
Department of Theoretical Physics, Irkutsk State University }\\[2cm]
\end{center}

\vspace*{0.5cm}

\begin{center}
{\Large \bf Abstract}
\end{center}
\vspace{0.5cm}

We study the effect of unitary mixing of scalar and vector fields
in general $\xi$ gauge. This effect takes place for non--conserved
vector currents and $\xi$ gauge generates some additional problems
with unphysical scalar field. We obtained solutions of
Dyson-Schwinger equations and perfomed the renormalization of full
propagators. The key feature of renormalization is the usage of
Ward identity, which relates some different Green functions. We
found that using Ward identity leads to disappearing of
$\xi$-dependence in renormalized matrix element.

\newpage
\setcounter{page}{1}

\section{Introduction}

The mixing of scalar and vector fields (S-V mixing)\ appears at the
loop level if the non-diagonal loop connecting
scalar and vector propagators exists.  This effect takes place when vector current is not conserved.

Similar effect was noticed before \cite{Bau,Aoki} when researching
the Standard Model in $\xi$ gauge where appears the mixing between
gauge boson field and unphysical field(so-called Higgs ghost) with
propagator pole at the point $p^2=\xi M^2$. \footnote{This
field has different names but we shall call it below just ghost.}
However physical scalar fields also can participate in mixing.
Thus, in \cite{K97} this effect in system $\pi - a_1$ was
considered, and in \cite{Pey} the S-V mixing between gauge bosons
and Higgs particles in extended electroweak models was
investigated. However in \cite{Pey} the problem of renormalization
which in this case is rather non-trivial, as well as the problem
of gauge dependency were not investigated. Note that consideration
of S-V mixing in $\xi$ gauge leads to interesting effect
\cite{Cheng}: the full propagators will have another type of
singularity then the bare ones. Namely, simple pole of bare
propagator at the point $p^2=\xi M^2$ turns into a double pole of
full propagator. After that the question of whether the Standard
Model is renormalizable in this gauge arises \cite{Cheng}.

In the present work we consider the unitary mixing of physical
scalar and physical vector in $\xi$ gauge. We focus on
renormalization of matrix element and its dependence on $\xi$. We
consider both boson and fermion loops that have got some special
features. Particulary our consideration is applicable for
electroweak models with extended Higgs sector.

$\xi$ gauge is defined by adding the following gauge fixing term to lagrangian
\begin{equation}
L_{gf} = -\frac{1}{2\xi}(\partial^{\mu}A_{\mu})^2.
\label{}
\end{equation}

The mixing of three bare propagators appears when taking into
account the loop contributions:\\
scalar particle propagator
\begin{eqnarray}
  \pi_{11} = \frac{1}{p^2-\mu ^2}\hspace{55mm} \parbox{40mm} {\includegraphics{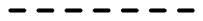}} ,
\label{p1}
\end{eqnarray}
vector field propagator in $\xi$ gauge
\begin{eqnarray}
  \pi_{22}^{\mu \nu} = \frac{1}{p^2- M ^2} \left\{ - g^{\mu \nu}+
p^\mu p^\nu
  \frac{(1-\xi )}{p^2-\xi M^2} \right\} \hspace{10mm}
  \parbox{50mm} {\includegraphics{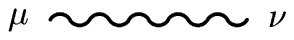}}
\label{p2}
\end{eqnarray}
and ghost propagator
\begin{eqnarray}
\pi_{33} = \frac{1}{p^2- \xi M ^2}\hspace{55mm} \parbox{40mm}
{\includegraphics{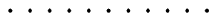}} . \label{pi33}
\end{eqnarray}
It is useful to divide vector propagator into transversal and
longitudinal parts
\begin{equation}
\pi_{22}^{\mu\nu}=
T^{\mu\nu}\cdot \frac{1}{p^2-M^2}+
L^{\mu\nu}\cdot \frac{\xi}{\xi M^2-p^2}, \ \ \ \ \ \
T^{\mu\nu}=-g^{\mu\nu}+\frac{p^\mu p^\nu}{p^2},\ \ \ \
L^{\mu\nu}=\frac{p^\mu p^\nu}{p^2}.
\label{}
\end{equation}
We should mention again that only longitudinal part of vector
propagator, which is not connected with spin J=1, can mix with
scalar fields. Vector current, with which the vector field
interacts, should not be conserving to acquire this effect.
In $\xi$ gauge the longitudinal part of propagator has the
unphysical pole at the point $p^2=\xi M^2$. However, it was noticed before
(see, e.g., \cite{Teylor}) that contribution of this pole in matrix element
has an opposite sign in
comparison with scalar meson contribution. Thus, for its
cancellation at tree level it is enough to add scalar field with
propagator of the kind (\ref{pi33}). At loop level there arises a
mixing of above mention propagators and, moreover, the full
non-diagonal propagators, which do not exist at tree level,
appear. So, first of all, we should study the problem of
renormalization of coupled propagators and S-matrix (in)dependence
on the gauge parameter.

\section{The Dyson-Schwinger equations system}

In case of mixing the propagators and loops aquire  matrix
structure and the Dyson-Schwinger equation takes the form
\footnote{As compared with consideration of unitary gauge
\cite{K97} we redefine non-diagonal propagators scalar-vector
$i\Pi_{12}\to\Pi_{12}$, $i\Pi_{21}\to\Pi_{21}$ for more symmetry.
Furthermore, we do not assume any symmetry for non-diagonal
transitions in advance - the symmetry relations are the
consequence of form of interaction and can vary. Note that such
changing of equations is just the redefinition of loops.}.
\begin{eqnarray}
\Pi_{11}&=&\pi_{11}-\Pi_{11}J_{11}\pi_{11}
-\Pi_{12}^\mu J_{21}^\mu \pi_{11}-\Pi_{13}J_{31}\pi_{11} \nonumber\\
\Pi_{12}^\mu&=&-\Pi_{11}J_{12}^\nu\pi_{22}^{\nu\mu}
-\Pi_{12}^\gamma J_{22}^{\gamma\nu}\pi_{22}^{\nu\mu}
-\Pi_{13}J_{32}^\nu\pi_{22}^{\nu\mu} \nonumber\\
\Pi_{13}&=&-\Pi_{11}J_{13}\pi_{33}
-\Pi_{12}^\mu J_{23}^\mu \pi_{33}-\Pi_{13}J_{33}\pi_{33} \nonumber\\
\Pi_{21}^\mu &=&-\Pi_{21}^\mu J_{11}\pi_{11}
-\Pi_{22}^{\mu\gamma} J_{21}^\gamma \pi_{11}
-\Pi_{23}^\mu J_{31}\pi_{11} \nonumber\\
\Pi_{22}^{\mu\nu}&=&\pi_{22}^{\mu\nu}
-\Pi_{21}^\mu J_{12}^\gamma \pi_{22}^{\gamma\nu}
-\Pi_{22}^{\mu\gamma} J_{22}^{\gamma\rho} \pi_{22}^{\rho\nu}
-\Pi_{23}^\mu J_{32}^\gamma \pi_{22}^{\gamma\nu} \nonumber\\
\Pi_{23}^\mu &=&-\Pi_{21}^\mu J_{13}\pi_{33}
-\Pi_{22}^{\mu\gamma} J_{23}^\gamma \pi_{33}
-\Pi_{23}^\mu J_{33}\pi_{33} \nonumber\\
\Pi_{31}&=&-\Pi_{31}J_{11}\pi_{11}
-\Pi_{32}^\mu J_{21}^\mu \pi_{11}-\Pi_{33}J_{31}\pi_{11} \nonumber\\
\Pi_{32}^\mu &=&-\Pi_{31}J_{12}^\nu \pi_{22}^{\nu\mu}
-\Pi_{32}^\gamma J_{22}^{\gamma\nu} \pi_{22}^{\nu\mu}
-\Pi_{33}J_{32}^\nu \pi_{22}^{\nu\mu} \nonumber\\
\Pi_{33}&=&\pi_{33}-\Pi_{31}J_{13}\pi_{33}
-\Pi_{32}^\mu J_{23}^\mu \pi_{33}-\Pi_{33}J_{33}\pi_{33}
\label{}
\end{eqnarray}
Here $\pi_{ij}$ are bare propagators, \ $\Pi_{ij}$ are full propagators, \
$J_{ij}$ are the one particle irredicible loop contributions.
In this equations values with two indexes need to be divided into
transversal and longitudinal parts.
\begin{eqnarray}
\Pi_{22}^{\mu\nu}&=& T^{\mu\nu}\Pi_{22}^T (p^2)+L^{\mu\nu}\Pi_{22}^L (p^2)
\nonumber \\
\pi_{22}^{\mu\nu}&=& T^{\mu\nu}\pi_{22}^T (p^2)+L^{\mu\nu}\pi_{22}^L (p^2)
\nonumber \\
J_{22}^{\mu\nu}&=& T^{\mu\nu}J_{22}^T (p^2)+L^{\mu\nu}J_{22}^L (p^2),
\label{}
\end{eqnarray}
where
\[
T^{\mu\nu}= -g^{\mu\nu}+\frac{p^\mu p^\nu}{p^2},\ \ \ \ \ \ \
L^{\mu\nu}=\frac{p^\mu p^\nu}{p^2}.
\]
In values with one index we shall come to scalar functions
according to
\begin{eqnarray}
\Pi_{12}^\mu(p)&=&p^\mu \Pi_{12}(p^2),\ \ \ \
\Pi_{21}^\mu(p)= p^\mu \Pi_{21}(p^2)  \nonumber \\
J_{12}^\mu(p)&=&p^\mu J_{12}(p^2),\ \ \ \ \
J_{21}^\mu(p)=p^\mu J_{21}(p^2)  \nonumber \\
\Pi_{23}^\mu(p)&=&p^\mu \Pi_{23}(p^2),\ \ \ \
\Pi_{32}^\mu(p)=p^\mu \Pi_{32}(p^2)  \nonumber \\
J_{23}^\mu(p)&=&p^\mu J_{23}(p^2),\ \ \ \ \
J_{32}^\mu(p)=p^\mu J_{32}(p^2).
\label{}
\end{eqnarray}
One can see that equation for the transversal component is
separates itself from the system and has the same form as in the
absence of scalar-vector mixing.
\begin{equation}
\Pi_{22}^T = \pi_{22}^T - \Pi_{22}^T J_{22}^T \pi_{22}^T,
\end{equation}
the solution of which is
\begin{equation}
\Pi_{22}^T = \frac{1}{p^2 - M^2 + J_{22}^T}.
\end{equation}

As for longitudinal components we have the following system of
equations in $\xi$ gauge
\footnote{Further we work with longitudinal components and drop index L except as otherwise noted.}
\begin{eqnarray}
\Pi_{11}&=&\pi_{11}-\Pi_{11}J_{11}\pi_{11}
-p^2 \Pi_{12} J_{21} \pi_{11}-\Pi_{13}J_{31}\pi_{11} \nonumber\\
\Pi_{12}&=&-\Pi_{11}J_{12}\pi_{22}
-\Pi_{12} J_{22} \pi_{22}
-\Pi_{13}J_{32}\pi_{22} \nonumber\\
\Pi_{21}&=&-\Pi_{21}J_{11}\pi_{11}
-\Pi_{22} J_{21} \pi_{11}
-\Pi_{23}J_{31}\pi_{11} \nonumber\\
\Pi_{22}&=&\pi_{22}
-p^2 \Pi_{21}J_{12}\pi_{22}
-\Pi_{22} J_{22} \pi_{22}
-p^2 \Pi_{23} J_{32} \pi_{22} \nonumber\\
\Pi_{13}&=&-\Pi_{11}J_{13}\pi_{33}
-p^2\Pi_{12} J_{23} \pi_{33}-\Pi_{13}J_{33}\pi_{33} \nonumber\\
\Pi_{31}&=&-\Pi_{31}J_{11}\pi_{11}
-p^2\Pi_{32} J_{21} \pi_{11}-\Pi_{33}J_{31}\pi_{11} \nonumber\\
\Pi_{23}&=&-\Pi_{21}J_{13} \pi_{33}
-\Pi_{22}J_{23} \pi_{33}
-\Pi_{23}J_{33}\pi_{33} \nonumber\\
\Pi_{32}&=&-\Pi_{31}J_{12} \pi_{22}
-\Pi_{32}J_{22} \pi_{22}
-\Pi_{33}J_{32}\pi_{22} \nonumber\\
\Pi_{33}&=&\pi_{33}-\Pi_{31}J_{13}\pi_{33}
-p^2 \Pi_{32} J_{23} \pi_{33}-\Pi_{33}J_{33}\pi_{33}.
\label{}
\end{eqnarray}
The solution of the system is
\begin{eqnarray}
\Pi_{11}&=&\frac{1}{D}\left[
(\pi_{22}^{-1}+J_{22})(\pi_{33}^{-1}+J_{33})-s J_{23} J_{32}
\right]  \nonumber \\
\Pi_{12}&=&-\frac{1}{D}\left[
J_{12}(\pi_{33}^{-1}+J_{33})- J_{13} J_{32}
\right]  \nonumber \\
\Pi_{21}&=&-\frac{1}{D}\left[
J_{21}(\pi_{33}^{-1}+J_{33})-J_{23}J_{31}
\right]  \nonumber \\
\Pi_{22}&=&\frac{1}{D}\left[
(\pi_{11}^{-1}+J_{11})(\pi_{33}^{-1}+J_{33})- J_{31} J_{13}
\right]  \nonumber \\
\Pi_{13}&=&-\frac{1}{D}\left[
J_{13}(\pi_{22}^{-1}+J_{22})-s J_{12}J_{23}
\right]  \nonumber \\
\Pi_{31}&=&-\frac{1}{D}\left[
J_{31}(\pi_{22}^{-1}+J_{22})-s J_{32}J_{21}
\right]  \nonumber \\
\Pi_{23}&=&-\frac{1}{D}\left[
J_{23}(\pi_{11}^{-1}+J_{11})-J_{21}J_{13}
\right]  \nonumber \\
\Pi_{32}&=&-\frac{1}{D}\left[
J_{32}(\pi_{11}^{-1}+J_{11})-J_{31}J_{12}
\right]  \nonumber \\
\Pi_{33}&=&\frac{1}{D}\left[
(\pi_{11}^{-1}+J_{11})(\pi_{22}^{-1}+J_{22})-s J_{21} J_{12}
\right],
\label{solutions_xi}
\end{eqnarray}
where $s=p^2$ and
\begin{eqnarray}
D(s)=(\pi_{11}^{-1}+J_{11})(\pi_{22}^{-1}+J_{22})(\pi_{33}^{-1}+J_{33})
-(\pi_{11}^{-1}+J_{11}) s J_{32} J_{23}- \nonumber \\
-(\pi_{22}^{-1}+J_{22}) J_{31}J_{13} -(\pi_{33}^{-1}+J_{33})s
J_{21} J_{12} + s J_{12}J_{23}J_{31} + s J_{32}J_{21}J_{13} .
\label{}
\end{eqnarray}
We should mention that transversal and longitudinal parts of
vector propagator are not fully independent. The condition
$J_{22}^T(0)+J_{22}^L(0)=0$ is necessary for the matrix element
not to have the pole $1/p^2$.

\section{Boson loops:\ $\pi - a_1$ system}

Here we consider the same model that was studied in \cite{K97}
in unitary gauge: $\pi - a_1$ system, which is dressed by the $\pi \sigma$ intermediate state.

Feynman rules for the given model are
\begin{eqnarray}
&&\includegraphics{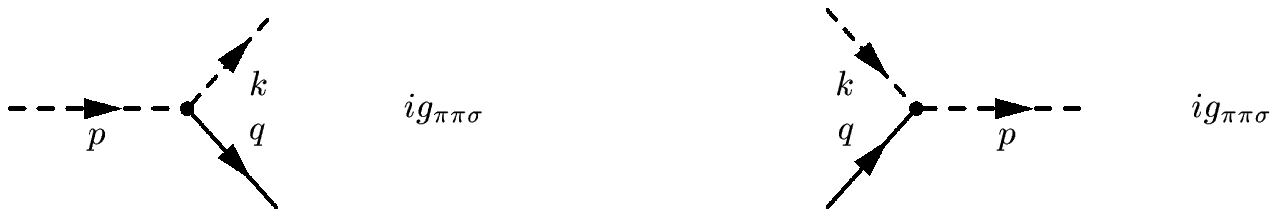} \nonumber\\
&&\includegraphics{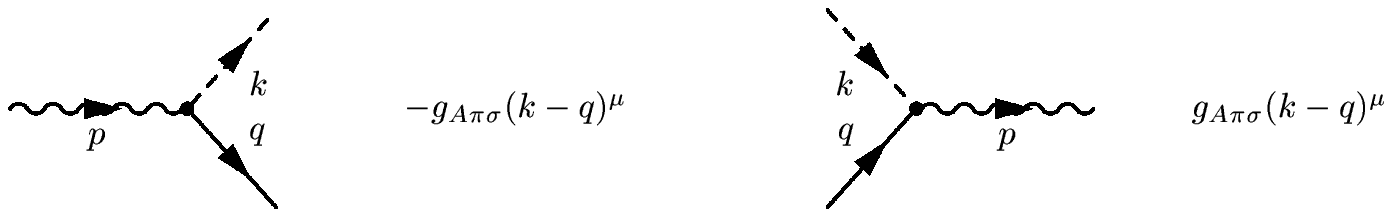} \nonumber\\
&&\includegraphics{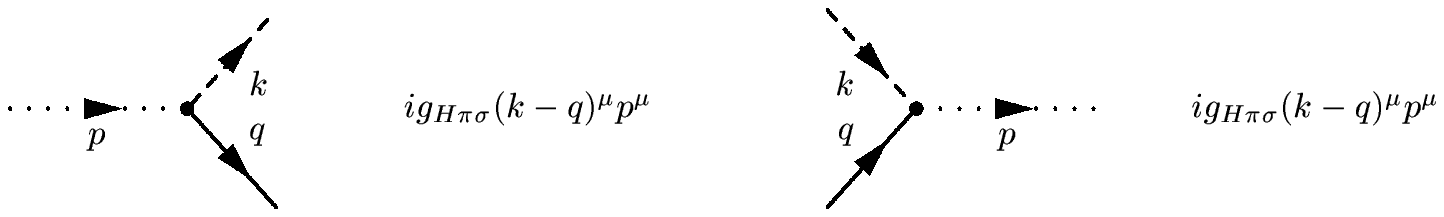}.\nonumber
\end{eqnarray}

To cancel unphysical poles in matrix element at tree level it is
necessary to fix the ghost coupling constant:
\begin{equation}
g_{G\pi\sigma}=\frac{g_{a\pi\sigma}}{M}.
\label{}
\end{equation}

Let us write down the result of calculations of loop contribution. Note that after
determining of Feynman rules the loops should be agreed upon Dyson-Schwinger equations.
We prefer to calculate loops using Landau-Cutkosky rules.\\
\begin{eqnarray}
J_{11}(p^2) &=& -i\ g_{\sigma\pi\pi}^2
\int \frac{d^4 l}{(2\pi)^4}\
\frac{1}{(l^2 - \mu^2)((l-p)^2 - m^2)}, \nonumber\\  \nonumber\\
J_{12}^{\mu}(p) &=& -\ g_{a_1\pi\sigma} \ g_{\sigma\pi\pi}
\int \frac{d^4 l}{(2\pi)^4}\
\frac{(2l-p)^{\mu}}{(l^2 - \mu^2)((l-p)^2 - m^2)}, \nonumber\\  \nonumber\\
J_{13}(p^2) &=& -i\frac{g_{\pi\pi\sigma}g_{a_1\pi\sigma}}{M}
\int \frac{d^4 l}{(2\pi)^4}
\frac{(p\cdot(2l-p))}{(l^2-\mu^2)((l-p)^2-m^2)}, \nonumber\\  \nonumber\\
J_{22}^{\mu\nu}(p) &=& -i\ g_{a_1\pi\sigma}^2
\int \frac{d^4 l}{(2\pi)^4}\
\frac{(2l-p)^{\mu} (2l-p)^{\nu}} {(l^2 - \mu^2)((l-p)^2 - m^2)}, \nonumber\\ \nonumber\\
J_{23}^\mu (p) &=& \frac{g_{a_1\pi\sigma}^2}{M}
\int \frac{d^4 l}{(2\pi)^4}
\frac{(2l-p)^\mu (p\cdot(2l-p))}{(l^2-\mu^2)((l-p)^2-m^2)}, \nonumber\\ \nonumber\\
J_{33}(p^2) &=&
-i\frac{g_{a_1\pi\sigma}^2}{M^2}
\int \frac{d^4 l}{(2\pi)^4}
\frac{(p\cdot(2l-p)) (p\cdot(2l-p))}{(l^2-\mu^2)((l-p)^2-m^2)}.
\label{loops}
\end{eqnarray}
Here $m=m_\sigma,\ \mu=m_\pi$.
Feynman rules lead to symmetry relations for non-diagonal loops.
\begin{eqnarray}
J_{21}^{\mu} &=& - J_{12}^{\mu},  \nonumber\\
J_{32}^\mu  &=& -J_{23}^\mu ,  \nonumber\\
J_{31} &=& J_{13}.
\end{eqnarray}

All loops are expressed in terms of one function $H(p^2)$ with
some subtractive polynomials which must be defined at
renormalization.
\footnote{Here and below we do indicate limits of integration in dispersion
integrals: they are from threshold to infinity.}
\begin{equation}
H(p^2) =
\frac{1}{\pi}\ \int \frac{ds}{s (s - p^2)}
\left( \frac{\lambda(s,m^2,\mu^2)}{s^2} \right)^{1/2}
\end{equation}
\begin{eqnarray}
J_{11}(s)&=&g_1^2[P_{11}+sH(s)]  \nonumber \\
J_{12}(s)&=& -i g_1 g_2[P_{12}+ H(s)]  \nonumber \\
J_{13}(s)&=&\frac{g_1 g_2}{M}[P_{13} + sH(s)]  \nonumber \\
J_{22}(s)&=&g_2^2[P_{22}+H(s)]  \nonumber \\
J_{23}(s)&=& i \frac{g_2^2}{M}[P_{23}+H(s)]  \nonumber \\
J_{33}(s)&=&\frac{g_2^2}{M^2}[P_{33}+sH(s)],
\label{loops2}
\end{eqnarray}
where $P_{ij}$ are  polynomials by \ s \ with real coefficients. We
introduced notations:\ $ g_1=g_{\sigma\pi\pi}/\sqrt{16 \pi},\
g_2=(\mu^2-m^2) g_{a_1\pi\sigma}/\sqrt{16 \pi}$,\ \
$\lambda(a,b,c)=(a-b-c)^2-4bc$.

The matrix element $\pi\sigma\to\pi\sigma$ with full propagators
has the form:
\begin{equation}
\frac{1}{16\pi}{\cal M}^{J=0}=-g_1^2 \Pi_{11}-2i g_1 g_2 \Pi_{12}
- 2\frac{g_1 g_2}{M} \Pi_{13} + 2i \frac{g_2^2}{M} \Pi_{23}
-\frac{g_2^2}{s} \Pi_{22}-\frac{g_2^2}{M^2} \Pi_{33}.
\label{Smatrix}
\end{equation}

\noindent \underline{Renormalization of pion pole}

We will use the renormalization scheme with subtraction on mass
shell. It is clear that procedure is more complicated due to
mixing of propagators. Requirements for renormalization of pion
pole can be formulated in the most simple way:
\begin{itemize}
\item Function D(s) has a simple zero at the point $s=\mu^2$ at
any values of coupling constants which are supposed to be
independent.
\item Full pion propagator $\Pi_{11}$ has pole with unit residue like the bare $\pi_{11}$.
It means that the sum of all loop insertion to external pion line
is equal to zero.
\end{itemize}
These requirements lead to conditions on loops at the point
$s=\mu^2$ i.e. on subtractive polynomials.
\begin{eqnarray}
J_{11}(\mu^2)&=&J_{11}^{\prime}(\mu^2)=0,  \nonumber \\
J_{12}(\mu^2)&=&0,  \nonumber \\
J_{13}(\mu^2)&=&0 .
\label{}
\end{eqnarray}

\noindent \underline{Renormalization of $\xi$}

As far as the mass of the vector particle is renormalized in
transversal part of vector propagator we can consider M as
renormalized mass. Thus, renormalization of unphysical pole at the
point $s=\xi M^2$ is renormalization of gauge parameter $\xi$.

Let us try to act by the analogy with pion pole and to formulate
renormalization requirements in a following way \footnote{These
requirements are minimal and mean that after the dressing
propagators have the same type of singularity. Since ghost appears
only in propagatobrs but not as external lines we do not set any
requirements to residues of propagators.}
\begin{itemize}
\item Function D(s) has a zero of second order at the point $s=\xi M^2$ at any
values of coupling constants.
\item  Full propagators $\Pi_{22}$ and $\Pi_{33}$ have simple pole at this point.
\end{itemize}
These reqirements with usage of solutions (\ref{solutions_xi})
will give the following conditions
\begin{equation}
J_{22}(\xi M^2)=J_{33}(\xi M^2)=J_{12}(\xi M^2)=
J_{13}(\xi M^2)=J_{23}(\xi M^2)=0.
\label{renxi}
\end{equation}

It is easy to see that among full propagators, besides $\Pi_{22}$
and $\Pi_{33}$, only $\Pi_{23}$ can have the pole at the point
$s=\xi M^2$. Therefore, it is enough to write down only these
contributions to trace the unphysical pole in the matrix element.
\begin{equation} \frac{1}{16\pi}\hat{\cal
M}^{J=0}=-\frac{g_2^2}{p^2}\Pi_{22}(p^2)
-\frac{g_2^2}{M^2}\Pi_{33}(p^2) + 2i \frac{g_2^2}{M}\Pi_{23}(p^2).
\label{}
\end{equation}
Using solutions of Dyson-Schwinger equatons and requirements (\ref{renxi}) we
can find the necessary condition for absense of unphysical pole in matrix element:
the function $Y(s)$
\begin{equation}
Y(s)=M^2 J_{33}(s)+s J_{22}(s)+ 2i Ms J_{23}(s)
\label{no_pole}
\end{equation}
have a second order zero at the point $\xi M^2$. This is
condition on subtractive polynomials since as we can see from
(\ref{loops2}) loops function $H(p^2)$ is cancellated in (\ref{no_pole}).

Let us recall that the absence of pole $1/p^2$ in matrix element relates $J_{22}^T$
and $J_{22}^L$ is
\begin{equation}
J_{22}^T(0) +  J_{22}^L(0) = 0 .
\label{no_pole2}
\end{equation}
With accounting (\ref{renxi}) polynomial in the loop
$J_{22}^L$ must have the following form
\begin{equation}
P_{22} = E\left( 1 - \frac{s}{\xi M^2} \right) - \frac{s}{\xi M^2} H(\xi M^2) ,
\end{equation}
where E is some fixed constant which is defined in the transversal part of loop $J_{22}$.
 Now it is possible to write out renormalized
loops satisfying the condition (\ref{no_pole}).
\begin{eqnarray}
J_{22} &=& \left[ E\left( 1 - \frac{s}{\xi M^2} \right) - \frac{s}{\xi M^2} H(\xi M^2)
+ H(s)   \right],  \nonumber \\  \nonumber \\
J_{23} &=& i \frac{g_2^2}{M} \left[ H(\xi M^2)- H(s)   \right] ,  \nonumber \\ \nonumber \\
J_{33} &=& \frac{g_2^2}{M^2} \left[ -\xi M^2 H(\xi M^2)-
E \xi M^2 \left( 1 - \frac{s}{\xi M^2} \right) + sH(s)   \right] .
\label{}
\end{eqnarray}
The rest of loops which do not take part in function Y(s)
(\ref{no_pole}):
\begin{eqnarray}
J_{11}&=& g_1^2 \left[ -sH(\mu^2) - \mu^2 H'(\mu^2)(s-\mu^2) + sH(s)
\right],  \nonumber \\  \nonumber \\
J_{12}&=& -ig_1 g_2 \left[ \frac{\xi M^2 H(\mu^2)-\mu^2 H(\xi M^2)}{\mu^2-\xi M^2} +
s \frac{H(\xi M^2) -  H(\mu^2)}{\mu^2-\xi M^2} + H(s) \right],  \nonumber \\ \nonumber \\
J_{13}&=& \frac{g_1 g_2}{M} \left[ \mu^2 \xi M^2
\frac{H(\xi M^2) -  H(\mu^2)}{\mu^2-\xi M^2}
- \frac{\xi M^2 H(\xi M^2)-\mu^2 H(\mu^2)}{\mu^2-\xi M^2}
- s H(s) \right].
\label{}
\end{eqnarray}

Now, after we defined subtractive polynomials in the loops, we can
calculate matrix element. Substituting full propagators we
obtain cumbersome expression which has evident dependence on gauge
parameter $\xi$. So we can conclude that renormalization of
unphysical pole by analogy with physical can not be done and must be realized in other way.

\noindent \underline{Ward identity.}

The key feature in renormalization is the usage of Ward identity,
which relates some different Green functions. It is obtained in
\cite{Cheng} and has the form
\begin{equation}
\langle 0 | T{(\partial_{\mu}A^{\mu}(x)-\xi M \varphi(x))
(\partial_{\nu}A^{\nu}(y)-\xi M \varphi(y))} |0\rangle =0,
\label{WI}
\end{equation}
where $\varphi(x)$ is ghost field.

Let us recall that Ward identity (\ref{WI}) was obtained in
\cite{Cheng} in the simpler case with the help of BRST
transformation. For any case we will obtain it in a
different way.

The above mentioned Feynman rules correspond to the following lagrangian
\footnote{We did not write here isotopic indexes because they were trivial for our model.}
\begin{eqnarray}
{\cal L}= -\frac{1}{4}(\partial_{\mu}A_{\nu}-\partial_{\nu}A_{\mu})^2
+\frac{1}{2}M^2 A_{\mu}A^{\mu}-\frac{1}{2\xi}(\partial A)^2
\nonumber \\
+\frac{1}{2}(\partial_{\mu}\varphi)^2
-\frac{1}{2}\xi M^2 \varphi^2+A_{\mu}J^{\mu}+
\frac{1}{M}\partial_{\mu}\varphi J^{\mu}.
\label{}
\end{eqnarray}
Here $J^\mu$ is vector current, we will not concretize it.
We wrote here only terms with vector and ghost fields. Motion equations have the form
\begin{eqnarray}
(\partial_{\alpha}\partial^{\alpha}+M^2)A_{\mu}
-(1-\frac{1}{\xi})\partial_{\mu}(\partial A)
&=&-J_{\mu}  \\
(\partial_{\alpha}\partial^{\alpha}+\xi M^2)\varphi
&=&-\frac{1}{M}(\partial J) .
\label{}
\end{eqnarray}
Consequence of these equations is
\begin{equation}
(\partial_{\alpha}\partial^{\alpha}+\xi M^2)
((\partial A)-\xi M \varphi)=0.
\label{}
\end{equation}
It means that appeared combination of fields $(\partial A)-\xi M
\varphi$ is non-interacting field. So the two point Green function
of this combination should not change under interactions. For the
case of bare propagators we have following expression
\begin{equation}
\langle 0 | T\left\{ ((\partial A(x))-\xi M \varphi(x))
((\partial A(y))-\xi M \varphi(y))\right\} |0\rangle=0.
\label{}
\end{equation}
To obtain it we must accurately differentiate T-product of vector
fields. In this procedure there appear additional terms
proportional to simultaneous commutators of interacting fields.
However, it is well known that simultaneous commutative relations
of interacting fields are coincided with the same for free fields, see e.g. \
cite{BS}.
After some calculations we have
\begin{equation}
\frac{\partial}{\partial x_{\mu}}\frac{\partial}{\partial y_{\nu}}
\langle T\{A_{\mu}(x)A_{\nu}(y)\}\rangle_0 =
\langle T\{\partial A(x)\partial A(y)\}\rangle_0
-i\xi\delta^4(x-y).
\label{}
\end{equation}
As a result, we come to Ward identity in terms of full propagators
\footnote{Note that there exists a disagreement on exact form of this relation.
In particular in \cite{Pass} it  was written without last term.}
\begin{equation}
s \Pi_{22}^L(s) -2i s\xi M \Pi_{23}(s)+\xi^2 M^2 \Pi_{33}(s)+\xi = 0.
\label{ch}
\end{equation}
When substituting explicit form of full propagators
(\ref{solutions_xi}) into the Ward identity (\ref{ch}) we shall
get some relations for the loop contributions. Note that we
consider the coupling constants $g_1$,$g_2$ as independent and (\ref{ch})
gives few conditions for loops.
\begin{eqnarray}
M^2 J_{33}+s J_{22}+2i s M J_{23}&=&0, \nonumber \\
J_{22}J_{33}+s(J_{23})^2&=&0,  \nonumber  \\
2is M J_{12}J_{13} - M^2 J_{13}^2 + s^2 J_{12}^2&=&0,  \nonumber  \\
-J_{22}J_{13}^2 + s J_{33}J_{12}^2 +2s J_{12}J_{13}J_{23}&=&0
\label{loops_xi}
\end{eqnarray}

Having resolved this equations we get the simple relations between
loops \footnote{It is useful to express all loops via $J_{22}$
because it has one more restriction from condition
(\ref{no_pole2}).}
\begin{eqnarray}
J_{33}=\frac{s}{M^2} J_{22},  \nonumber \\
J_{23}=\frac{i}{M} J_{22},  \nonumber \\
J_{13}= i \frac{s}{M}J_{12}.
\label{relat}
\end{eqnarray}
We notice that the same function Y(s) (\ref{no_pole}) appears in
(\ref{loops_xi}) which guarantees the absence of unphysical pole
in the matrix element.

Now we calculate the full propagators using relation (\ref{loops_xi})
following from Ward identity. One can see at once that in the function $D(s)$
dependence on gauge parameter is factorized.
\footnote{We note that usually under dressing the pole
lying above threshold shifts to complex plane. But in our case
under Ward identities the pole stays at real axis.}

\begin{equation}
D(s)=-\frac{(s-\xi M^2)^2}{\xi M^2} \hat{D}(s),
\end{equation}
where appears function
\begin{equation}
\hat{D}(s)=(\pi_{11}^{-1}+J_{11})(M^2+J_{22})+s(J_{12})^2,
\label{}
\end{equation}
playing the same role in unitary gauge \cite{K97}.

The full propagators acquire very simple form
\begin{eqnarray}
\Pi_{11}&=&\frac{(M^2+J_{22})}{\hat{D}}  \nonumber \\
\Pi_{12}&=&\frac{\xi M^2 J_{12}}{(s-\xi M^2)\hat{D}}  \nonumber \\
\Pi_{13}&=& -i\frac{M s J_{12}}{(s-\xi M^2)\hat{D}}  \nonumber \\
\Pi_{23}&=& i \frac{\xi M \left [(\pi_{11}^{-1}+J_{11})J_{22}+s(J_{12}^2)\right ]}
{(s-\xi M^2)^2\hat{D}}  \nonumber \\
\Pi_{22}&=&-\xi \frac{\left[(\pi_{11}^{-1}+J_{11})
(M^2 (s-\xi M^2)+s J_{22})+s^2(J_{12})^2\right]}
{(s-\xi M^2)^2\hat{D}}  \nonumber \\
\Pi_{33}&=& M^2\frac{\left [(\pi_{11}^{-1}+J_{11})
((s-\xi M^2)-\xi J_{22})-\xi s(J_{12})^2 \right ]}
{(s-\xi M^2)^2\hat{D}}.
\label{propr}
\end{eqnarray}

When we substitute the full propagators into the matrix element
$\pi\sigma\to\pi\sigma$ (\ref{Smatrix}) we shall find
\begin{equation}
\frac{1}{16\pi}{\cal M}^{J=0}= -g_1^2\frac{(M^2+J_{22})}{\hat{D}}+
2i g_1 g_2\frac{J_{12}}{\hat{D}}-
g_2^2\frac{(\pi_{11}^{-1}+J_{11})}{\hat{D}}.
\label{}
\end{equation}
We see that dependence on gauge parameter $\xi$ has disappeared
and this expression coincides with matrix element in unitary gauge
if there are no conditions on loops $J_{11}$, $J_{12}$, $J_{22}$
at the point $s=\xi M^2$.

Finally if we group the terms in the matrix element (\ref{Smatrix})
in the following way
\begin{equation}
\frac{1}{16\pi}{\cal M}^{J=0}= -g_1^2\Pi_{11} -
2ig_1 g_2 (\Pi_{12}-\frac{i}{M} \Pi_{13}) -
g_2^2 (\frac{1}{s}\Pi_{22}+\frac{1}{M^2} \Pi_{33} - \frac{2i}{M} \Pi_{23}),
\label{Smat1}
\end{equation}
we find that not only sum but each of this three addends do not
depend on gauge parameter $\xi$.

\section{Fermion loops: W,Z -- Higgs mixing in extended electroweak models}

Unitary mixing between gauge bosons and Higgs particles is
possible only in the extended electroweak models since
pseudoscalar or charged Higgs particles are required for that.
In Standard Model where exists only one scalar Higgs this effect is absent.
We do not define concretely the model but just fix form of vertex.

\noindent \underline{Mixing $W^{\pm}$ -- scalar Higgs.}

Interaction vertexes have the form
\begin{eqnarray}
&&\includegraphics{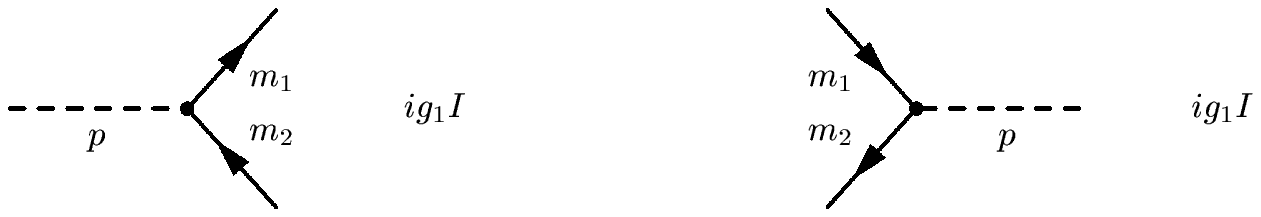} \nonumber\\
&&\includegraphics{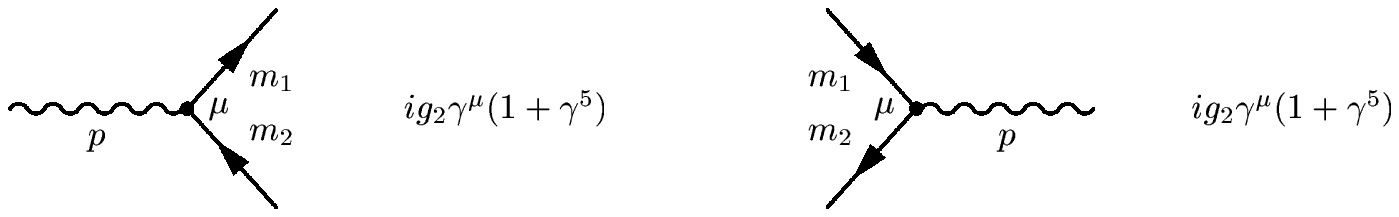} \nonumber\\
&&\includegraphics{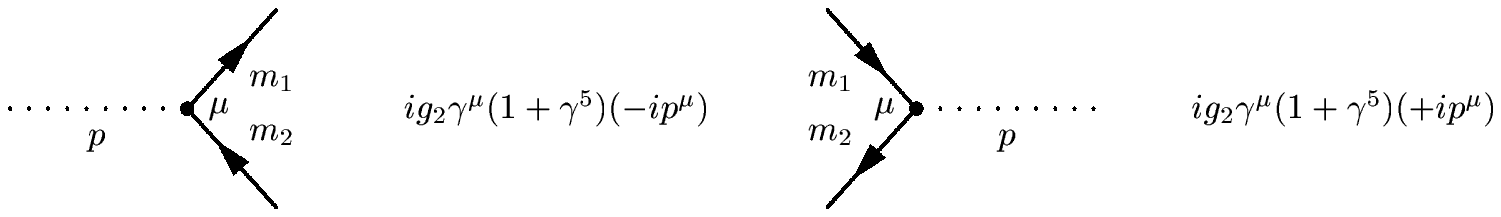}\nonumber
\end{eqnarray}
Loops take the form
\begin{eqnarray}
  J_{11}(p^2) &=& -ig_{1}^2 \int \frac {d^4 l}{(2 \pi )^4}
  Sp\left\{I \frac {1} {\hat{l}-\hat{p}-m_2} I  \frac {1} {\hat{l}-m_1} \right\}
  \nonumber \\ \nonumber \\
  J_{12}^\mu(p)&=& -ig_{1}g_{2} \int \frac {d^4 l}{(2 \pi )^4}
  Sp\left\{I \frac {1} {\hat{l}-\hat{p}-m_2} \gamma ^\mu(1+\gamma^5)
  \frac {1} {\hat{l}-m_1} \right\}
  \nonumber \\ \nonumber \\
  J_{13}(p^2)&=&\frac{g_{1}g_{2}}{M} \int \frac {d^4 l}{(2 \pi )^4}
  Sp\left\{I \frac {1} {\hat{l}-\hat{p}-m_2} \hat{p}(1+\gamma^5)
  \frac {1} {\hat{l}-m_1} \right\}
  \nonumber \\ \nonumber \\
  J_{22}^{\mu \nu}(p)&=& -ig_{2}^2 \int \frac {d^4 l}{(2 \pi )^4}
  Sp\left\{\gamma ^\mu (1+\gamma^5) \frac {1} {\hat{l}-\hat{p}-m_2}
  \gamma ^\nu (1+\gamma^5)  \frac {1} {\hat{l}-m_1} \right\}
  \nonumber \\ \nonumber \\
  J_{23}^\mu(p)&=& \frac{g_{2}^2}{M} \int \frac {d^4 l}{(2 \pi )^4}
  Sp\left\{\gamma ^\mu (1+\gamma^5) \frac {1} {\hat{l}-\hat{p}-m_2} \hat{p}
  (1+\gamma^5)  \frac {1} {\hat{l}-m_1} \right\}
  \nonumber \\ \nonumber \\
  J_{33}(p)&=& -\frac{g_{2}^2}{M^2} \int \frac {d^4 l}{(2 \pi )^4}
  Sp\left\{\hat{p}(1+\gamma^5) \frac {1} {\hat{l}-\hat{p}-m_2} \hat{p}
  (1+\gamma^5)  \frac {1} {\hat{l}-m_1} \right\}.
\end{eqnarray}

Symmetry properties become rather different as compared with
boson loops.
\begin{eqnarray}
  J_{21}^\mu(p)&=& J_{12}^\mu(p)\nonumber \\
  J_{31}(p)&=& -J_{13}(p)\label{Jsymmetry} \\
  J_{32}^\mu(p)&=& - J_{23}^\mu(p)\nonumber
\end{eqnarray}

In the fermion case all longitudinal loops are expressed in terms of the two functions
$H_1(p^2)$, $H_2(p^2)$ with some subtractive polynomials.
\begin{eqnarray}
 H_1(p^2) &=& \frac{1}{\pi}\ \int \frac{(m_1+m_2)^2-s}{s (s - p^2)}
 \left(\frac{\lambda(s,m^2,\mu^2)}{s^2} \right)^{1/2}ds\nonumber \\
 H_2(p^2) &=& \frac{p^2}{\pi}\ \int
 \frac{(m_1-m_2)^2-s(m_1^2+m_2^2)}{s^2 (s - p^2)}
 \left(\frac{\lambda(s,m^2,\mu^2)}{s^2} \right)^{1/2}ds \nonumber
\end{eqnarray}
\begin{eqnarray}
 J_{11}=f_1^2\left[P_{11}+sH_1(s)\right] &,& J_{12}=f_1f_2\left[P_{12}+H_1(s)\right]\nonumber \\
 J_{13}=i\frac{f_1f_2}{M}\left[P_{13}+sH_1(s)\right] &,& J_{22}=\hat{f}_2^2\left[P_{22}+
 H_2(s)\right] \nonumber \\
 J_{23}=i\frac{\hat{f}_2^2}{M}\left[P_{23}+H_2(s)\right]&,& J_{33}=\frac{\hat{f}_2^2}{M^2}
 \left[P_{33}+sH_2(s)\right]
\label{loops3}
\end{eqnarray}
where $P_{ij}$ are polynomials by \ s \ with real coefficients.
Notations are $f_1=g_1/\sqrt{8 \pi}$, $f_2=(m_1^2 - m_2^2) g_2/
\sqrt{8\pi}$, $\hat{f}_2 = g_2/ \sqrt{4 \pi}$.

If to look at relation (\ref{relat}) following from Ward identity
it is easy to see that Ward identity puts constraints only on
subtractive polynomials, as for loop integrals $H_1$, $H_2$ they
identically satisfy (\ref{relat}). If these relations are
satisfied we obtain simple $\xi$ dependence on  (\ref{propr}) in
propagators and we only need to trace the $\xi$ dependence in the
matrix element.

Matrix element $f_1(q_1) \overline{f}_2(q_2)\to f_1(k_1)
\overline{f}_2(k_2)$ have form
\begin{eqnarray}
 {\cal M}^{J=0}&=&- g_{1}^2 \Pi_{11}\ \overline{v}(q_2)u(q_1)\cdot \overline{u}(k_1)v(k_2)-
 \nonumber \\
 &&- g_{1} g_{2} \left( \Pi_{12}-\frac{i}{M}\Pi_{13}\right)\ \overline{v}(q_2)u(q_1)\cdot
 \overline{u}(k_1)\hat{p}(1+\gamma^5)v(k_2)-\nonumber \\
 &&- g_{1} g_{2} \left( \Pi_{21}+\frac{i}{M}\Pi_{31}\right)\ \overline{v}(q_2)\hat{p}
 (1+\gamma^5)u(q_1)\cdot\overline{u}(k_1)v(k_2)- \\
 &&- g_{2}^2\left(\frac{1}{s}\Pi_{22}+\frac{1}{M^2} \Pi_{33}-\frac{i}{M}\Pi_{23}+
 \frac{i}{M}\Pi_{32}\right)\ \overline{v}(q_2)\hat{p}(1+\gamma^5)u(q_1)\cdot\overline{u}(k_1)
 \hat{p}(1+\gamma^5)v(k_2).    \nonumber
 \label{matrixel}
\end{eqnarray}
It is possible to simplify this expression using motion equation for spinors but it is
clear that different spinor matrix
elements in (\ref{matrixel}) are accompanied by $\xi$ independent factors (see (\ref{Smat1})).
Thus, the dependence on gauge parameter in matrix element disappears.

\noindent \underline{Mixing $W^{\pm}(Z^0)$ -- pseudoscalar Higgs}

Vertexes of interaction Higgs with ferimons have form
\begin{eqnarray}
&&\includegraphics{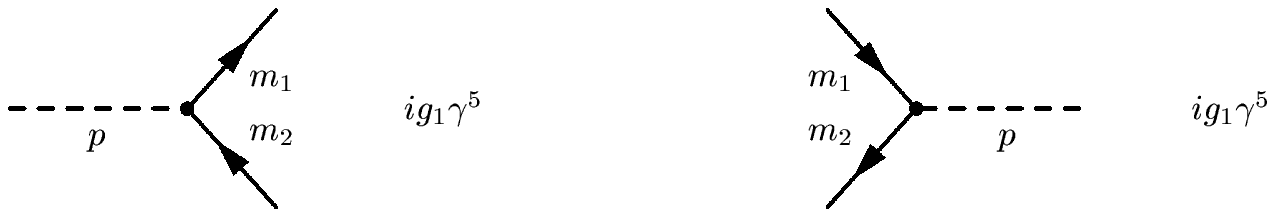} . \nonumber
\end{eqnarray}

The matrix element $f_1(q_1) f_2(q_2)\to f_1(k_1) f_2(k_2)$
slightly changes
\begin{eqnarray}
 {\cal M}^{J=0}&=& -g_{1}^2 \Pi_{11}\overline{v}(q_2)\gamma^5 u(q_1)\cdot\overline{u}(k_1)\gamma^5 v(k_2)
 -\nonumber \\
 &&-g_{1} g_{2} \left( \Pi_{12}-\frac{i}{M}\Pi_{13}\right)\overline{v}(q_2)\gamma^5 u(q_1)
 \cdot\overline{u}(k_1)\hat{p}(1+\gamma^5)v(k_2)-\nonumber \\
 &&-g_{1} g_{2} \left( \Pi_{21}+\frac{i}{M}\Pi_{31}\right)\overline{v}(q_2)\hat{p}(1+\gamma^5)u(q_1)
 \cdot\overline{u}(k_1)\gamma^5v(k_2)-\\
 &&-g_{2}^2\left(\frac{1}{s}\Pi_{22}+\frac{1}{M^2} \Pi_{33}-\frac{i}{M}\Pi_{23}+
 \frac{i}{M}\Pi_{32}\right)\ \overline{v}(q_2)\hat{p}(1+\gamma^5)u(q_1)\cdot\overline{u}(k_1)
 \hat{p}(1+\gamma^5)v(k_2).    \nonumber
\label{matrixel_psedo}
\end{eqnarray}
Some loops have changed in comparison with scalar Higgs.
\begin{eqnarray}
  J_{11}(p^2) &=& -ig_{1}^2 \int \frac {d^4 l}{(2 \pi )^4}
  Sp\left\{\gamma^5 \frac {1} {\hat{l}-\hat{p}-m_2} \gamma^5
  \frac {1} {\hat{l}-m_1} \right\}
   \nonumber \\ \nonumber \\
  J_{12}^\mu(p)&=& -ig_{1}g_{2} \int \frac {d^4 l}{(2 \pi )^4}
  Sp\left\{\gamma^5 \frac {1} {\hat{l}-\hat{p}-m_2} \gamma ^\mu(1+\gamma^5)
  \frac {1} {\hat{l}-m_1} \right\}
   \nonumber \\  \nonumber \\
  J_{13}(p^2)&=&\frac{g_{1}g_{2}}{M} \int \frac {d^4 l}{(2 \pi )^4}
  Sp\left\{\gamma^5 \frac {1} {\hat{l}-\hat{p}-m_2} \hat{p}(1+\gamma^5)
  \frac {1} {\hat{l}-m_1} \right\}
\end{eqnarray}
\begin{equation}
  J_{21}^\mu(p)=J_{12}^\mu(p),\ \ \
  J_{31}(p)= -J_{13}(p)
\label{Jsymmetry_psedo}
\end{equation}
After calculating these loops one can see that only difference with scalar Higgs is
another form of the function $H_1$.
\begin{eqnarray}
 H_1(p^2) = \frac{1}{\pi}\ \int \frac{(m_1-m_2)^2-s}{s (s - p^2)}
 \left(\frac{\lambda(s,m^2,\mu^2)}{s^2} \right)^{1/2}ds\nonumber
\end{eqnarray}
\begin{eqnarray}
 J_{11}&=&f_1^2\left[P_{11}+sH_1(s)\right] ,
 J_{12}=f_1f_2\left[P_{12}+H_1(s)\right]\nonumber \\
 J_{13}&=&i\frac{f_1f_2}{M}\left[P_{13}+sH_1(s)\right] \nonumber
\label{loops_psedo}
\end{eqnarray}
Here we use another notation for $f_2$ :
$f_2^2=(m_1 + m_2)^2 g_2^2/8 \pi$. \\
Renormalized matrix element
 $f_1(q_1) f_2(q_2)\to f_1(k_1) f_2(k_2)$  (\ref{matrixel}) takes the form
\begin{eqnarray}
 {\cal M}^{J=0}&=&-g_{1}^2 \frac{(M^2+J_{22})}{\hat{D}} \overline{v}(q_2)\gamma^5u(q_1)
 \cdot\overline{u}(k_1)\gamma^5v(k_2)+\nonumber \\
 &&+2g_{1} g_{2} \frac{J_{12}}{\hat{D}}
 \left(\overline{v}(q_2)\gamma^5u(q_1)\cdot\overline{u}(k_1)\hat{p}(1+\gamma^5)v(k_2)+
 \overline{v}(q_2)\hat{p}(1+\gamma^5)u(q_1)\cdot\overline{u}(k_1)\gamma^5v(k_2)\right)-\nonumber\\
 &&-g_2^2\frac{(\pi_{11}^{-1}+J_{11})}{\hat{D}} \overline{v}(q_2)\hat{p}(1+\gamma^5)u(q_1)
 \cdot\overline{u}(k_1) \hat{p}(1+\gamma^5)v(k_2).
\end{eqnarray}

One can see that dependence on $\xi$ in matrix element  has been disapeared.

\section{Summary}

We investigated the effect of unitary mixing scalar-vector in general
$\xi$ gauge and found that under usage of Ward identity the
renormalized matrix element does not depend on gauge parameter.
The interesting feature noted in \cite{Cheng} in simpler case
consist in changing of singularity type after dressing. Simple
pole $1/(p^2-\xi M^2)$ in bare propagators after dressing turns into double pole.
Such possibility always exists in
mixing of two bare propagators with same masses but it is realized
only at definite relations between loops which follow from the
Ward identity.

In \cite{Cheng} the boson loop contributions were calculated and
it was found that position of double unphysical pole is diverged
under the usage of Ward identity. It is resulted in the opinion
that Standard Model \footnote{In \cite{Cheng} it was investigated
mixing "longitudinal part W --- ghost" in the Standard Model, that
is partial case of our consideration.} is not renormalizable in
$\xi$ gauge.

We see from above that Ward identity leads to simple relations
between loops (\ref{relat}) and if these relations are fullfiled the position of
double pole is ultraviolet stable. So we can suppose that in
calculations of \cite{Cheng} the obtained loops do not satisfy the
Ward identity although this identity is used in general form.

After our investigation it seems that the usage of $\xi$ gauge for
extended Higgs model is not convinient. But this gauge(and its
particulary cases) is widely spread in investigations of electroweak
models. Particulary it is possible to control the correctness of
calculations varying $\xi$ and tracing the variation in the matrix
element.

The physical consequences of unitary mixing "scalar---vector" in extended electroweak
models deserves further investigation.

We are intended to V.V.Lyubushkin for verification of some formulae.


\end{document}